
\magnification=1200
\def\ni{\noindent}
\def\.{\mathaccent 95}
\def\a{\alpha}
\def\be{\beta}

\def\la{\lambda}

\def\De{\Delta}

\def\frac#1#2{{\textstyle{{#1}\over {#2}}}}
\def\ni{\noindent}
\def\lsim{\mathrel{\rlap{\lower4pt\hbox{\hskip1pt$\sim$}}
    \raise1pt\hbox{$<$}}}
\def\gsim{\mathrel{\rlap{\lower4pt\hbox{\hskip1pt$\sim$}}
    \raise1pt\hbox{$>$}}}
\def\sqr#1#2{{\vcenter{\vbox{\hrule height.#2pt
         \hbox{\vrule width.#2pt height#1pt \kern#1pt
         \vrule width.#2pt}
         \hrule height.#2pt}}}}

% Next 5 lines define \lapprox and \gapprox: "less than or approximately
% equal to" and "greater than or approximately equal to".
\newbox\grsign \setbox\grsign=\hbox{$>$} \newdimen\grdimen \grdimen=\ht\grsign
\newbox\simlessbox \newbox\simgreatbox
\setbox\simgreatbox=\hbox{\raise.5ex\hbox{$>$}\llap
     {\lower.5ex\hbox{$\sim$}}}\ht1=\grdimen\dp1=0pt
\setbox\simlessbox=\hbox{\raise.5ex\hbox{$<$}\llap
     {\lower.5ex\hbox{$\sim$}}}\ht2=\grdimen\dp2=0pt

% 
% from Larry Molnar
% Set up some definitions:  
%
%This is how to have an approximate sign under < or > :

\def\ref#1  {\noindent \hangindent=24.0pt \hangafter=1 {#1} \par}
\def\doublespace {\smallskipamount=6pt plus2pt minus2pt
                  \medskipamount=12pt plus4pt minus4pt
                  \bigskipamount=24pt plus8pt minus8pt
                  \normalbaselineskip=24pt plus0pt minus0pt
                  \normallineskip=2pt
                  \normallineskiplimit=0pt
                  \jot=6pt
                  {\def\smallskip {\vskip\smallskipamount}}
                  {\def\medskip   {\vskip\medskipamount}}
                  {\def\bigskip   {\vskip\bigskipamount}}
                  {\setbox\strutbox=\hbox{\vrule 
                    height17.0pt depth7.0pt width 0pt}}
                  \parskip 12.0pt
                  \normalbaselines}
\def\ts{\times}
\def\lb{\langle}
\def\rb{\rangle}

\def\bfvp{{\bf v}'}
\def\bfjp{{\bf j}'}

\def\bfwp{{\bomega}'}

\def\b0{b'^{(0)}}
\def\v0{v'^{(0)}}
\def\w0{\omega'^{(0)}}
\def\bb0{\bfbp^{(0)}}
\def\bv0{\bfvp^{(0)}}
\def\bw0{\bfwp^{(0)}}
\def\bj0{\bfjp^{(0)}}

\doublespace
%\twocolumn[\hsize\textwidth\columnwidth\hsize\csname @twocolumnfalse\endcsname

\centerline{\bf Variability Associated with $\a$ Accretion Disc Theory} 
\centerline{\bf for Standard and Advection Dominated Discs}
\centerline{Eric G. Blackman}
\centerline{Institute of Astronomy, Madingley Road, Cambridge CB3 OHA, 
 England}              
\medskip
\centerline{\it submitted to MNRAS} 
%\medskip
\centerline {\bf ABSTRACT}

The $\alpha$ turbulent viscosity formalism for accretion discs 
must be interpreted as a mean field theory. The extent to which the disc 
scale exceeds that of the turbulence determines the precision of the predicted luminosity $L_\nu$. The assumption of 
turbulence and use of $\a$ implies: (1) Field line stretching generates a 
magnetic pressure  $\gsim \a^2/6$ of the total pressure generally,
and a 1 to 1 relation between $\a$ and the pressure ratio when 
shearing instabilities dominate the viscosity. (2)
Large eddy sizes and speeds in typical 
advection dominated accretion flows (ADAFs)
lead to a lower precision in  $L_\nu$ 
than for thin discs of a given total observation duration and central mass. The allowed 
variability (relative precision) 
at a particular frequency increases (decreases) with the size of the 
contributing region. For X-ray binary type ADAFs, the allowed variability is 
$\sim 5$\% at $R \le 1000$ Schwarzchild radii for averages over
$\gsim 1000$sec. But for large galactic nuclei like  
NGC 4258 and M87, the relative precision error can approach $50-100\%$  
even at $R \le 100 R_S$ for currently available observation durations.
More data are then required to compare with ADAF predictions.

%The possibility that ADAFs might not be turbulent, but
%instead might transport angular momentum by a wind 
%needs to be investigated, in which case the constraints
%could be relaxed.

\ni Key Words: accretion discs; galaxies: nuclei, active; 
Galaxy: centre; turbulence; binaries: general

\vfill
\eject

\ni{\bf 1. Introduction }

Accretion discs are a widely accepted paradigm (e.g Pringle 1981; 
Papaloizou \& Lin 1995) to explain  a variety of  features in high energy 
astrophysical sources such as active galactic nuclei (AGN), X-ray binary 
systems, cataclysmic variables (CVs), and dwarf novae.
As accreting gas orbits a central massive source, internal energy dissipation  
drains the rotational energy, allowing material to move in and
angular momentum out.  The dissipation sustains steady accretion and some 
fraction of the dissipated energy accounts for  the observed luminosity.  
Micro-physical viscosities are too small to explain observed luminosities 
so an enhanced transport mechanism, likely involving turbulence,  is essential.
Since astrophysical discs are surely magnetized to some non-zero 
level, the ``Balbus-Hawley'' shearing instability (e.g. Balbus \& Hawley 1991; 
c.f. Balbus \& Hawley 1998), which produces self-sustaining turbulence,
is a natural and likely ubiquitous driver of angular momentum transport
for at least thin discs and possibly thick discs as well.
Significant dissipation may also occur above a thin disc, in a 
corona (e.g Haardt \& Maraschi 1993; Field \& Rogers 1993; DiMatteo et al. 
1997). For low enough accretion rates, the dissipated energy may be
primarily advected rather than radiated (Ichimaru 1977; Rees et al., 1982; 
Narayan \& Yi 1995ab) forming an advection dominated
accretion flow (ADAF) thick disc.  ADAFs have been effective in modeling 
quiescent phases of accretion in a variety of X-ray binaries and
galactic nuclei (see Narayan et al. 1998a, for a review).

While non-linear instabilities in thin discs 
have been extensively simulated locally 
(e.g. Brandenburg et al 1995; Stone et al., 1996 Balbus et al., 1996) 
a useful approach to global disc 
models has been to swipe the details of the stress tensor
into a turbulent viscosity of the form  (Shakura 1973; Shakura \& Sunyaev 1973)
$$\nu_{tb}=\a c_s H\simeq v_{tb}l_{tb},\eqno(1)$$ 
where $H$ is the disc height, $c_s$ is the 
sound speed, $l_{tb}$ is the dominant turbulent eddy scale, 
$v_{tb}$ is the eddy speed at that scale, 
and $\a< 1$ is taken to be a constant.  Use of this formalism 
{\it requires} a mean field theory.  Viscous coupling of differentially 
rotating fluid elements is a local paradigm, 
so assumptions of azimuthal symmetry
and steady radial inflow (e.g. Pringle 1981; Narayan \& Yi 1995ab) 
require turbulent motions to be statistically smoothed
over the time and/or spatial scales on which mean quantities vary.

The required mean field approach is similar to
that employed in mean-field magnetic dynamo theory (Parker 1979), 
where the induction equation for the magnetic field is averaged and solved.
In the kinematic (and therefore incomplete) dynamo theory, the velocity
is imposed and the momentum equation is ignored.  
For the simplest global $\a$ accretion disc approach, 
the focus is reversed (but also incomplete); the momentum, energy and 
continuity equations are solved, with the inclusion of the magnetic field
as a pressure rather than employing the magnetic induction equation
(e.g. Narayan \& Yi 1995ab).  However, the usual disc equations 
are not usually derived from the formal
averaging approach.  Balbus \& Hawley (1998) have addressed some 
of these points, but the conditions for which  
the standard equations result when $H\sim R$ are not 
commonly studied.

Despite being incomplete, the  $\a$ formalism provides a useful framework
for thin and ADAF discs. 
The absence of much scale separation between $H$ and $R$ 
has important consequences. 
Here I show that $l_{tb}\le H$ and field line stretching 
place a non-trivial lower limit on the magnetic
energy for large $\a$.  A 1-to-1 relation between
$\a$ and $\beta$ results when the viscosity is due to shearing instabilities
as shown in section 7.  
I also estimate the precision of the $\a$ turbulent 
disc formalism and interpret this for thin and thick ADAF discs.
For ADAFs the precision is lower, and thus the allowed variability 
higher than for thin discs with  a given central mass, and fixed 
observation duration.  A discussion of the implications for stellar vs. 
galactic nuclei presumed ADAF systems is given. 

Balbus \& Gammie (1994) present a nice 
discussion of fluctuations in a thin disc and 
a relation to luminosity, but a different approach and
different questions are addressed here for both thin and thick discs.
Although quantities like velocity can always 
be formally separated into mean and fluctuating parts, the turbulence 
gives a negligible non-zero RMS error to the mean only
when the disc radius is much larger than the scale of the turbulence.
This RMS error will be estimated here as a 
precision measure of the $\a$ formalism, as a function of the averaging time.

\ni{\bf 2. Precision of $\a$-Accretion Disc Theory}

The usual slim disc equations are derived (e.g. Abramowicz et al. 1988;
 Narayan \& Yi 1995a) by 
vertically averaging the continuity, Navier-Stokes, and energy equations 
and with the magnetic field incorporated only as an additional pressure. 
%$$\partial_R(\rho R H v_R) = 0 \eqno()$$
%$$v_R\partial_Rv_R-\Om^2R+\Om_k^2R+\rho^{-1}\partial_R(\rho c_s^2)=0\eqno()$$
%$$v_R\partial_R(R^2\Om)-(\rho RH)^{-1}\partial_R(\a \rho c_s^2R^3H/\Om_k)\part%ial_R\Om=0\eqno()$$
%$$\Sigma v_R T\partial_Rs-3(1+\ep)\rho H v_R\partial_Rc_s^2+2c_s^2Hv_R\partial%_R \rho =(2\a f\rho c_s^2R^2H/\Omega_k)(\partial_R\Om)^2.\eqno()$$
%This omission is 
%the complement of what is done in kinematic magnetic dynamo theory,
%where there the magnetic induction equation is solved without consideration
%of the momentum equation.  
%The averaging is usually done only vertically, 
%and assumed to be formally 
%be derived from averaging over azimuth as well.  
%A more rigorous approach would be to write mean and 
%fluctuating components for each quantity, 
%derive separate equations for the mean 
%and fluctuation components,
%%(e.g. letting the velocity ${\bf v}={\bar {\bf v}}+{\bf v}_{tb}$)
%and calculate the stress tensor from these fluctuations
%(see Balbus 1998, Balbus \& Gammie 1994). 
%It is not necessarily true that the standard $\a$ disc equations result.
Without presenting the formalism, here I assume
the equations of e.g. Narayan \& Yi (1995a) hold, 
but emphasize that the standard simple replacement of the micro-physical
viscosity with a turbulent viscosity hides the requirement
of radially and/or temporally averaging  
(in addition to the usual azimuthal and
vertical smoothing).  For the radial average, 
a scale $\xi$ must be chosen such that 
$l_{tb} < \xi < R,$ 
where $R$ is the disc radius.  The spatial average of a quantity like velocity ${\bf V}(R)$ 
is then  
${\bf V}_0(R) =\lb{\bf V}(R)\rb_s \simeq 
\lb{\bf V}(R,t')\rb_\xi= \int_{\la \le \xi} {\bf V}(R+\la,t') d\la,$
where the similarity follows from the assumption that
the time dependence is only due to turbulent fluctuations which
are intended to be smoothed for mean quantities.
The subscript $0$ indicates the mean quantity
to be used in standard $\a$ disc theory.
%The condition for validity can then also
%written
%$$t_{tb}< R/v_{tb}.\eqno(4)$$
%The $\xi$ determines the radial resolution of the theory.  
For a temporal average, taken over a duration $t_{obs}$, we have
${\bf V}_0(R)=
\lb{\bf V}(R)\rb_{t_{obs}}=(1/t_{obs})\int_t^{t+t_{obs}} {\bf V}(R,t' ) dt'.$
The temporal average is meaningful only over times such that $t_{obs}>t_{tb}$
where $t_{tb}$ is the dominant energy containing eddy turnover time
scale. 

How precise is the assumption that the mean 
speed behaves as a steady monotonic function of $R$ in the presence of 
turbulence?  Note that ``precision'' is taken to mean that 
defined in Bevington (1969) with the $\a$ disc theory as the measuring 
device. A relative precision error (RPE) measures
how effective the theory is at predicting, not how accurate the predictions 
are.  The RPE error around the total mean speed $V_0$ 
is given by 
$\De V_0/V_0=[(\De V_0)_{fl}/V_0 + \partial_R V_0(\De R)_{rs}/V_0]$,
and the two terms on the right measure two RMS contributions
to the RPE.  
%Assuming the turbulence is roughly isotropic for simplicity, 
%(which is not strictly accurate, Balbus \& Gammie 1994),
The first can be approximated by 
$(\De V_0)_{fl}=v_{tb}/N_{fl}^{1/2}\simeq v_{tb}(\xi/l_{tb} + t_{obs}/t_{tb})^{-1/2}
$ where $N_{fl}$ measures the 
``effective'' number of eddies per radial averaging length. With increasing 
$t_{obs}$, $N_{fl}$ can well exceed the ``snapshot'' $t_{obs}=0$ 
value  $\xi/l_{tb}$. 
The second RMS contribution to the RPE above results from the fact that 
$R-(\xi/2)N_{rs}^{-1/2}\le R \le R+ (\xi/2)N_{rs}^{-1/2}$ 
is indistinguishable once  $\xi$ is chosen.  
Here $N_{rs}^{-1/2}\simeq (1 + t_{obs}/t_{tb})^{-1/2}$ and measures
the ``effective'' number of averaging scales per $\xi$ 
which increases from its snapshot value of 1 for long $t_{obs}$.
The fluctuation and resolution numbers ($N_{fl}$ and $N_{rs}$) 
increase with $t_{obs}$ because the turbulence does not generate 
eddies in exactly the same location over time, so there is a  
smoothing effect which reduces the ``effective'' averaging and eddy scales. 

Since the total speed for both thin and thick discs
is dominated by a contribution $\propto R^{-1/2}$, 
we can then estimate a total RPE for $R$ as 
$$\De R/R\simeq [ 2(\De {V_0})_{fl}/{V_0}+(\De R)_{rs}/R ]
\simeq  [2(v_{tb}/{V_0})N_{fl}^{-1/2}+(\xi/2)N_{rs}^{-1/2}].\eqno(2)$$
Assuming a constant accretion rate, 
(2) translates into an RPE in the luminosity given by
$$\De L_\nu/(L_\nu) \simeq |\psi|(\De R/R)\eqno(3)$$ 
where $\nu$ is the frequency of emission and 
$|\psi|\equiv |R \partial_R [{\rm Ln}(L_\nu)]|$ which is usually 
$0.5 \le |\psi|\le 10$ as will be addressed later.
The RPE can be used to estimate the variability allowed for a given
$t_{obs}$.

Though phenomenologically derived, the RPE formulae have properties 
which show that they capture the limiting cases correctly:  
First, for $t_{obs}>>t_{tb}$, they are reduced as expected.  Second, 
for $t_{obs}=0$, there is an optimal scale of 
$$\xi/R=\xi_{opt}/R=(v_{tb}/V_{0})^{2/3}(l_{tb}/R)^{1/3} \eqno(4)$$
for which the error is minimized:  a larger 
$\xi$ reduces the RMS effect of the turbulent velocity, but one pays the price 
with a coarser spatial resolution.  
When $\xi_{opt}< l_{tb}$, 
the RPE is dominated by the resolution term but
then $\xi=l_{tb}$ must be used since it is the minimum allowed. 
This will  make  the RPEs relatively
independent of $\beta$, since changing $\beta$ changes the relative
importance of the two RPE terms while keeping (1) constant.

%This will be the case for thin discs. 
%When $\xi_{opt}\gsim l_{tb}$ the first term for even $t_{obs}=0$
%is also important and the RPE is larger. This will be the case for ADAFs.

{\bf 3. Energy Constraints and Relations Between Characteristic Speeds}

In a highly conducting 
turbulent plasma, the magnetic field is naturally amplified 
to the extent that $v_{tb}\simeq B/(4\pi \rho)^{1/2}\equiv v_A$,  the Alfv\'en speed (e.g. Parker 1979).  
Shearing box simulations, where turbulence is driven by a seed magnetic field  
(Stone et al. 1996; Brandenburg et al. 1995; Balbus \& Hawley 
1998), show $v_A\gsim v_{tb}$. Because of field line stretching, 
equipartition of turbulent and Alfv\'en 
speeds is generally a more applicable rule of thumb than 
any relation between the particle and magnetic pressures.

When the magnetic field is tangled on scales much smaller than 
those on which mean quantities vary
(which for ADAFs likely requires temporal averaging, as seen below)
averaging the Lorentz force gives an effective magnetic pressure
$$P_{mag}/\rho\simeq B^2/(24\pi\rho) =v_A^2/6=(1-\beta)\rho c_s^2,\eqno(5)$$
where $\beta$ is a parameter. Using (5) in (1) and $v_{tb}\simeq v_A$ 
we have 
$$l_{tb}=\a H/K_1^{1/2},\eqno(6)$$
where $K_1\equiv 6(1-\beta)$.
Because $l_{tb}\le H$, we have the constraint 
$$0 \le  \be \le (1-\a^2/6).\eqno(7)$$
The above relations also imply 
$$v_{tb}=K_1^{1/2}c_s. \eqno(8)$$

\ni {\bf 4. RPE  of Thin Disc Models}

For thin discs, $H<<R$ and 
%The longest possible eddy turnover time  
%satisfies  $t_{tb}\sim l_{tb}/v_{tb}\lsim H/\a c_s\sim \Om_k/\a << R/V_R$,
%where $R/V_R$ is the radial in-fall time. 
%For simplicity let us choose $\xi=R$.
$V_0 \simeq V_{0,\phi}$, the Keplerian speed.
From (3), (6) and (8)
% and using $v_{tb}/V_{0,\phi}=\a H/R$, 
we have 
$$\De L_\nu/L_\nu =|\psi|\De R/R \simeq 
2|\psi|(v_{tb}/V_{0,\phi})N_{fl}^{-1/2}+0.5|\psi|(\xi/R)N_{rs}^{-1/2}$$
$$\simeq {{2 |\psi| (H/R)(K_1/1)^{1/2}}\over
{\{{\rm Max}[(1,22({K_1\over 1})^{2/3}({\a\over 0.01})^{-2/3}]
+2.3\ts 10^{5}t_{obs}({K_1\over 1})({\a \over 0.01})^{-1}
({M \over M_\odot})^{-1}({R \over 10R_s})^{-3/2}\}^{1/2}}}$$
$$+{0.005|\psi|(\a/0.01)(K_1/1)^{-1/2}(H/R)
{\rm Max}[1,22(K_1/1)^{2/3}(\a/0.01)^{-2/3}]
\over {[1+
2.3\ts 10^{5}t_{obs}(K_1/1)(\a/0.01)^{-1}(M/M_\odot)^{-1}(R/20R_s)^{-3/2}]^{1/2}}}
,\eqno(9)$$
%$$\De L_\nu/L_\nu =|\psi|\De R/R \sim 
%2|\psi|(v_{tb}/V_{0,\phi})N_{ff}^{-1/2}+0.5|\psi|(\xi/R)N_{rs}^{-1/2}$$
%$$\sim 2 |\psi| \a (H/R) [\xi/H+2.3\ts 10^{5}t_{obs}(\a/0.01)^{-1}
%(M/M_\odot)^{-1}(R/10R_s)^{-3/2}]^{-1/2}$$
%$$+0.5|\psi|(\xi/R)[1+
%2.3\ts 10^{5}t_{obs}(\a/0.01)^{-1}(M/M_\odot)^{-1}(R/20R_s)^{-3/2}]^{-1/2}
%,\eqno(4)$$
where $\xi$ has been replaced by the Max[,]
as per the discussion below (4), and relations (6), (8) and 
the definition of $N_{fl}$ have been used.
%$l_{tb}$ because
%$\xi_{opt}< l_{tb}$ for thin discs as per the discussion above.
%%(v_{tb}/v_{tot})(R/l_{tb}+t_{obs}/t_{tb})^{-1/2}
%,\eqno()$$ 
The temperature in an optically thick thin disc 
goes as $T_s\propto R^{-3/4}$ (e.g. Frank, King \& Raine 1992). 
Then, for example, in the Rayleigh-Jeans regime $(h\nu<<kT(R))$ 
where 
$L_\nu \propto \nu^2\int_{R_{min}}^{R_{max}}T_e(r)rdr$, 
the luminosity within a radius $R$ 
at a given frequency goes as $L_\nu \propto \nu^2 R^{5/4}$.
Thus $|\psi|=5/4$.

The RPE of (9) is small compared to what will be found for ADAFs.  A 
careful check, keeping (7) in mind, ensures that for all allowed $\beta$
the RPE $\lsim |\psi|H/R$. (In section 7, we show that $\a$ and 
$\beta$ can be related.)
%$< |\psi| \a (H/R)^{3/2}$, which is usually $<<1$:
%The predicted precision error in the luminosity from () is then 
%$\De (L_\nu)/L_\nu < \a (5/4)(H/R)^{3/2},$ 
%which is $< 0.04\a $ for $H/R < 1/10$.  For intermediate frequencies, the
%luminosity curve traces the peak of successive
%black-bodies at decreasing radii, so $\nu L_\nu\propto \nu^3 \sim \nu_c^3
%\propto T^3\propto R^{-9/4}$.  The RPE is 
%therefore $< (9/4)\a(H/R)^{3/2}$ even for a snapshot.
The RPE is further reduced for large $t_{obs}$.
The low RPE results because $H/R<<1$ and $v_{tb}<<v_{0}\sim v_{0,\phi}$
for thin discs.

\ni {\bf 5. Implications and RPE for Thick ADAF Discs}

For thick ADAF discs things are more subtle.
From (7) and the standard ADAF choice of $\a=0.3$ (Narayan 1998a)
we have $0\le \beta \le 0.985$. 
Note when $l_{tb}=H$ the fact that $H\sim R$ makes 
this upper limit extreme. 
Defining $K_2\equiv 2/(7-2\beta)$ 
and using $c_s=K_2^{1/2}V_{ff}\sim (H/R)V_{ff}$ (Narayan et al. 1998a),
where $V_{ff}=(GM/R)^{1/2}$ is the free-fall speed,  
we have $v_{tb}=(K_1 K_2)^{1/2}V_{ff}$, and thus  from (6) and (8)
$$l_{tb}/v_{tb}=
t_{tb}=\a H/(K_1 c_s)=\a H/(K_1 K_2)^{1/2} V_{ff}=\a R/(K_1 V_{ff}).
\eqno(10)$$
Furthermore, defining $K_3\equiv  12(1-\beta)/(7-2\beta)$, 
the total mean speed is (Narayan 1998a)
$V_{0}\sim [(9\a^2 /4)K_2^2+K_3]^{1/2}V_{ff}$,
so we have  
%$$ v_{tb}/V_{0}\equiv K_4 \sim (K_1K_2)^{1/2}/[(9\a^2 /4)K_2^2+K_3]^{1/2}
%=12(1-\be)(7-2\be)/[9\a^2+(4-2\be)(7-2\be)].\eqno(11)$$
$$ K_4\equiv v_{tb}/V_{0}= (K_1K_2)^{1/2}/[(9\a^2 /4)K_2^2+K_3]^{1/2}
=[12(1-\be)(7-2\be)]^{1/2}/[9\a^2+12(1-\be)(7-2\be)]^{1/2}.\eqno(11)$$
This is $\sim 1$ over the allowed range of $0\le \beta \le 0.985$.
Using (3) (4), (10) and (11) 
%and $K_\ga=(c_s/V_0)\sim 
%[(\ga -1)^2\a^2/(\ga -5/9)^2 + 0.7(5/3-\ga)/(\ga-5/9)]^{1/2}$ 
%(Narayan 1998a) we then have
%larger than that of thin discs.
gives 
$$\De L_\nu/L_\nu =|\psi|\De R/R \sim 
2|\psi|(v_{tb}/V_{0})N_{fl}^{-1/2}+|\psi|(\xi/2)N_{rs}^{-1/2}$$
$$\simeq {{1.22|\psi|(K_4/1)(\a/0.3)^{1/2}}\over 
{\{{\rm Max}[1,3.9({\a \over 0.3})^{-{2\over 3}}({K_4 \over 1})^{2\over 3}({K_1\over 3})^{1\over 3}({K_2\over 0.58})^{-{1\over 3}}]
%5.2(\xi_{opt}\over R)(K_1\over 3)^{1\over 2}(K_2\over 0.58)^{-1 \over 2}
+2.4\ts 10^3 t_{obs}({K_1\over 3})
({M\over M_\odot})^{-1}({R\over 20R_s})^{-{3 \over 2}}\}^{-{1\over 2}}}}
$$
$$+{{0.06|\psi|(\a/0.3)(K_1/3)^{-{1\over 2}}(K_2/0.58)^{1\over 2}
{\rm Max}[1,3.9(\a/0.3)^{-{2\over 3}}(K_4/1)^{2\over 3}(K_1/3)^{1\over 3}(K_2/0.58)^{-{1\over 3}}]}\over {
[1+8\ts 10^3 t_{obs}(\a/0.3)^{-{1 \over 2}}(K_1/0.3)(M/M_\odot)^{-1}(R/20R_s)^
{-{3\over 2}}]^{-{1 \over 2}}}}.\eqno(12)$$
%(\xi_{opt}/R) 
Recall that $K_{1,2,3,4}$ all depend
only on $\beta$. The rigorous form for  
$\xi_{opt}$ from (4) has also been employed.  Over   
the allowed range $0\le \beta \le 1-\a^2/6$, 
the RPE is relatively insensitive to 
$\beta$ (though in section 7 we show that 
$\a$ and $\beta$ may be related.)
This is because decreasing $\beta$ increases
$l_{tb}$ while lowering $v_{tb}$ and vice versa. 
The RPE is sensitive to {\it both}  $v_{tb}$ and $l_{tb}$. 

%8\ts 10^3 t_{obs}(\a/0.3)^{-}
%(M/M_\odot)^{-1}(R/20R_s)^{-3/2}]^{-1/2}.\eqno(6)$$ 
%2 |\psi| \a (H/R) [\xi/H+2.3\ts 10^{3}\a^{-1}
%t_{obs}(M/M_\odot)^{-1}(R/10R_s)^{-3/2}]^{-1/2}+0.5|\psi|(\xi/R)[1+

%$$\De v_{tot}/v_{tot} \sim \De v_{ff}/v_{ff} \sim 
%\pm v_{tb}/(v_{ff}N^{1/2}) 
%\sim 0.5 N^{-1/2}\sim 0.3(\a/0.3)^{1/2} (2+t_{obs}v_{ff}/R)^{-1/2}$$
%$$\sim 0.3(\a/0.3)^{1/2}
%[2+8\ts 10^2t_{obs}(M/M_\odot)^{-1}(R/20R_s)^{-3/2}]^{-1/2},\eqno()$$
%where $R_s \equiv 2GM/R$, and $M$ is the central mass. 
%Thus from (),
%$$\De (\nu L_\nu)/(\nu L_\nu) =
%\pm 2|\psi|\De v_{ff}/v_{ff}\sim \pm |\psi|0.6(\a/0.3)^{1/2}
%[2+8\ts 10^2t_{obs}(M/M_\odot)^{-1}(R/20R_s)^{-3/2}]^{-1/2},
%\eqno()$$
%\sim (3.3 (\a/0.3)^{-1}+
%0.3(\a/0.3) t_{obs}v_{ff}/R)^{-1/2}$$
%$$\sim (3.3(\a/0.3)^{-1}+0.3(\a/0.3) t_{obs}/[4\ts 10^{-4}\a(M/M_\odot)(R/10R/%s)^{3/2}])^{-1/2},\eqno()$$

I now estimate $|\psi|$ for various emission regimes based on 
ADAF scaling relations (Mahadevan 1997).
Consider the radio Rayleigh-Jeans radio regime.
Here $L_\nu \sim L_{\nu_c}$
%\propto T_e^{21/5}\nu_c^{2/5}$ 
%\propto \nu_c^2\int^R_{R_{min}} T_e(r)rdr,$
where $\nu_c$ is the peak frequency at each $R<R_{max}$ and 
is determined by synchrotron absorption
In the ADAF, $\nu_c \propto B T_e^2\propto T_e^2 R^{-5/4}$,
and is therefore a function of $R$. The 
spectrum traces the envelope of peak frequencies, with each frequency
corresponding to a particular $R$.
For moderate accretion rates by ADAF standards (but 
below the critical value required for an ADAF solution)
%e.g. appropriate for  NGC 4258, 
compressive electron heating is not
important (Narayan et al. 1998a) and $\partial_RT_e=0$. 
Using $\nu=\nu_c$ we have 
$L_\nu\propto T_e^5 R^{-1/2}$ 
and $|\psi|=1/2$. 
%This also provides a lower limit to $|\psi|$ 
%for models with $T_e=T_e(R)$ because $T_e(R)$ decreases with $R$.
When  compressive heating is important, $\partial_RT_e(R)\ne 0$.  
Fitting the steadily falling $T_e(R)$ curve of Narayan et al., (1998b) 
I obtain $Log T_e(R)\sim 9.8-0.3Log (R/R_s)-0.06 [Log( R/R_s)]^2$, so then 
$$R\partial_R[Ln\ T_e(R)]=-0.7-0.3 Log (R/R_S),\eqno(13)$$
and $|\psi|=1/2+3.5 + 1.5 Log (R/R_s)$. 

For $\nu$ below $\nu_c=\nu_{c,min}$, the critical frequency
corresponding to the maximum disc radius $R_{max}$, 
the spectrum is simply $\propto \nu^2$ at fixed 
$R_{max}$, and in the constant $T_e$ regime, 
$|\psi|=2$.  For the $\partial_RT_e(R)\ne 0$
regime using (13),  $|\psi| =1.3 + 0.3 Log (R/R_s)$.

In the Compton dominated sub-millimeter/X-ray regime 
$L_\nu\propto \nu_c^{\a_c}T_e^5 R^{-1/2}\nu^{-\a_c}\propto 
T_e^{5+2\a_c} R^{-(2+5\a_c)/4} \nu^{-\a_c}$,
which is sensitive to the Comptonization 
parameter $\a_c$ and $|\psi|=0.5+5\a_c/4$ for constant $T_e(R)$.
In the regime where (13) is applicable, 
$|\psi|=0.5+ 5\a_c/4+(5+2\a_c)[0.7+0.3 Log(R/R_s)]$.
In the Bremsstrahlung dominated sub-millimeter X-ray regime,
$L_\nu\propto Ln (R_{max}/R_{min})F(T_e)T_e^{-1}Exp[-h\nu/kT_e]$,
where $F(T_e)$ is dominated by a term $\propto T_e$ in the
when $kT_e>m_ec^2$ and dominated by a term $\propto T_e^{1/2}$ 
when $kT_e<m_ec^2$.
If  $\partial_R T_e=0$, $|\psi|\sim 1/Ln(R_{max}/R_{min})\sim 0.4$, 
for $R_{max}/R_{min}=1000$, but 
this is sensitive to radial dependences of
$T_e$ since the latter appears in the exponential for 
the Bremsstrahlung regime.
In the limit that $T_e > mc^2$, 
$|\psi|\sim 2.2(\nu/10^{20} Hz)(T_e/3\ts 10^9 K)^{-1}[0.7+0.3 Log(R/R_s)]$.  
For $T_e(R)<mc^2$, $|\psi|\sim |[0.5-2.2(\nu/10^{20} Hz)(T_e/3\ts 10^9 K)^{-1}][0.7+0.3 Log(R/R_s)]|$.

Overall, for most regimes and reasonable $(R/R_s)$,  
$1/2 \le |\psi| \le 10$.

\ni {\bf 6. Implications for Observations of Presumed ADAFs}

We can see from (12) that for $M\sim 10M_\odot$, (e.g. X-ray binary
type systems) and $|\psi|\le 10$,  
predictions probing the the inner $20R_s$ and 
averages over $t_{obs}\gsim 10^3$sec 
are quite precise, that is $\De L_\nu/L_\nu\le 0.05$.
At $R=1000R_s$, and $|\psi|=1/2$, $\De L_\nu/L_\nu \sim 0.05$.
The allowed variability decreases as $t_{obs}^{-1/2}$.

Now consider the galactic nucleus of NGC 4258
with central mass $M\simeq 3.5 \ts 10^7 M_\odot$, 
For the radio emitting regime of this source near 22GHz,
Herrnstein et al. (1998) found no
detection of 22GHz emission in NGC 4258
with a 3-$\sigma$ upper limit of 220$\mu Jy$.
This frequency is safely in the Rayleigh-Jeans regime and
the best fitting models of ADAFs to NGC 4258 have $T_e(R)$
approximately constant in this regime.
Herrnstein et al. interpret this non-detection to mean that any ADAF
proposed for this source (Lasota et al 1996)
cannot extend outside radius defined by $\nu_c=22$GHz, namely
$R\sim 100R_s$. For the observations, $t_{obs}\sim 10^5$. 
But at 100$R_s$  
the $t_{obs}$ term does not contribute to significantly to (12).
Since in this regime $|\psi|\sim 1/2 $, we have from (12)
$\De L_\nu/L_\nu \sim 0.4$, so this would
reduce the significance of non-detection at 22 GHz to $\sim 1 \sigma$.

%Furthermore, the above was derived
%essentially with $H/R\sim 0.$. If $H=R$, then the RPE 
%would be $0.86$.  
%more data are needed to fully rule out an ADAF outside $100R_s$.

For the Galactic centre, the presumed central mass is $\simeq 2.5 \ts 10^6
M_\odot$. Then from (12), at $R=20R_s$, $t_{obs}$ must be 
$> 10^3$ seconds to contribute to significantly 
reducing the RPE. The X-ray observations and many radio observations 
above 10 GHz
when taken together provide enough total $t_{obs}$ 
for low RPE  in this range (Narayan et al. 1998b and refs. therein) 
However for frequencies $\le 1GHz$, 
$R\ge 1000R_s$, and the total $t_{obs}$ must be $\gsim 10^6$sec,
for which there is insufficient data. 
Using $|\psi|=1.65$ in (12), $\De L_\nu/L_\nu \gsim 1$.

Application of ADAFs to larger galactic nuclei (Fabian \& Rees 1995) 
require much longer observation times for precise predictions.
For M87, $M\sim 3\ts 10^9M_\odot$ so at 20$R_s$, the
required $t_{obs}$ time would be $\gsim 10^6$sec for the
$t_{obs}$ term in (12) to reduce the RPE well below 1, 
while at $1000R_s$ the the limit would be $\gsim  10^8$ seconds of total
time. X-ray observations have been made for $1.4 \ts 10^4$ seconds (Reynolds et
al., 1996) and 
radio observations have been made for only of order hours at particular
frequencies, e.g. $2\ts 10^4$sec at 1.7GHz (Reid et al., 1989).
and $7.2 \ts 10^3$sec at 22 GHz (Spencer \& Juror, 1986)
More data are definitely needed in these sources.

%The precision error from () is then 
%$\De(\nu L_\nu)/(\nu L_\nu)\lsim 0.7$,
%corresponding to a change by $+0.2$ or $-0.5$ on
%a $Log(\nu L_\nu)$ plot.

%so $\De (\nu L_\nu)/\nu L_\nu \lsim 0.8$,
%which corresponds to a change by $+0.3$ or $-0.7$ 
%on a $Log(\nu L_\nu)$ plot.  

%do so just to get a feel for why 
%the fact that they are large is so important:
%Comparing  ADAF models to the Galactic center in the low frequency
%radio range, we can see from Narayan et al 1998,
%that the range of $\De \nu L_\nu\sim +0.3$ or $-0.7$ might 
%directly account for any disagreement with the observations
%straight away. 

\ni {\bf 7. Conclusions}

The presumption that accretion discs are turbulent 
implies that the standard disc equations must be interpreted as mean 
field equations. 
Field line stretching equilibrates the turbulent and Alfv\'en speeds
so that the constraint on the ratio of magnetic to total pressures
(Eq. 7) arises once
it is assumed that the turbulent scale cannot exceed the disc
height.  For large $\a$ ADAFs this constraint is non-trivial and
requires $\beta \le 0.985$.  
For ADAFs, $H/R \sim 1$, so that $0.985$ is really an extreme upper limit.

%For large $\a$ thick discs, self-consistency of the
%$\a$ approach for ADAFs requires 
%near equipartition between the magnetic, thermal, turbulent,
%and bulk kinetic energies.  This results from the fact that steady solutions 
%require small turbulent eddy scales compared to the disc radius.

The large turbulent scales and/or speeds for ADAFs 
lead to a  relative precision error (RPE) or allowed variability 
in the predicted luminosity. Indeed, there are two contributions 
to the RPE.  The first results from velocity fluctuations 
and the second comes the fact that the theory is unresolved 
on scales below the radial scale over which the averaging is taken. 
RPEs are larger ADAFs  than for thin discs.
The RPEs of (9) and (12) can be used to roughly predict allowed   
deviations from the $\a$ theory $L_\nu$ for a given $t_{obs}$,
and indicate when longer observations are 
needed to properly compare with disc models. 
The total RPE is reduced over large $t_{obs}$ 
because such averaging amounts to smoothing over an ensemble of 
many turbulent realizations of the disc. 

The conclusions about the location of any
transition radius to an ADAF in NGC 4258 (Herrnstein et al. 1998) 
based on a 22GHz non-detection must be interpreted with this in mind
since (12) suggests that an ADAF can significantly over-predict the
luminosity. More observation time is then needed to 
fully rule out an ADAF outside 100$R_S$.  
For the Galactic centre below
1GHz, and for larger mass systems like M87 or M60 generally, more data 
than currently available are also needed to compare with 
ADAFs, since the RPEs only reduce significantly from 1 over long $t_{obs}$.

Finally, note that (9) and (12) do not necessarily 
presume that the Balbus-Hawley type (e.g. Balbus \& Hawley 1998) 
shearing instability is the only source of viscosity.  Making that assumption 
however, 
provides a 1-to-1 link between $\a$ and $\beta$: we have  in the steady-state 
$t_{tb}\sim$ the instability growth rate, so 
$t_{tb}\simeq R/v_{0,\phi}$.  Also, $v_A\sim v_{tb}/3^{1/2}$
from shearing simulations as discussed in section 3.  Thus 
%im v_{tb}\sim v_A R/3v_{0,\phi}$.
$\nu=\a c_s H\simeq v_{tb}l_{tb}\sim v_A^2(R/3v_{0,\phi})$ which implies that 
$\a \sim 2(1-\beta)$ for thin discs and 
$\a \sim 2(1-\beta)(K_2/K_3)^{1/2}=[2(1-\beta)/3]^{1/2}$ for 
%check
ADAFs. Eqns (9) and (12) are then sensitive to $\beta$  mainly through the
explicit $\a$.  
%As disussed below (9) and (12), they are otherwise
%insensitive to $\beta$.

%It could be that ADAF discs are not less turbulent
%and instead transport their angular momentum primarily by
%a wind. Could this be true for all ADAFs ?(!) 
%If so, the $\a$ formalism need not be 
%interpreted as turbulence and the RPEs
%could be lower. The wind could be ``ejection'' dominated 
%(e.g. Xu \& Chen 1997) so as not to produce much radiation
%This needs to be investigated.  If for example,
%ADAFs are largely pressure supported, 
%the Balbus-Hawley instability may be less effective in 
%generating turbulence.

%Observed variations within the RPE of (4) and (6)
%mean that the data are consistent with the variations
%expected of the theory from turbulence.  
%The longer the $t_{obs}$ the lower the RPE.
%When variations are observed in excess of the RPE for a
%given $t_{obs}$ additional or alternative physics is necessary.

---------------------

Acknowledgments: Thanks to T. DiMatteo, R. Mahadevan, \& U. Torkelsson 
for discussions.

\ni Abramowicz, M.A., et al, 1988, ApJ 332 646.

\ni Balbus, S.A. \& Hawley, J.F., 1998, Rev. Mod. Phys, 70, 1.

\ni Balbus, S.A., Gammie, C.F., Hawley, J.F., 1994, MNRAS, 271, 197.

\ni Balbus., S.A., Hawley, J.F., \& Stone J.A., 1996, ApJ, 467, 76.

\ni Bevington P.R., {\it Data Reduction and Error Analysis for the Physical Sciences}, 1969, (New York:  McGraw-Hill), p3.

\ni Brandenburg A. et al., 1995, 446 741. 

\ni DiMatteo, T., Blackman E.G. \& Fabian, A.C., 1997 MNRAS 291, L23.

%\ni DiMatteo, T., personal communication.

\ni Fabian, A.C. \&  Rees M.J., 1995, MNRAS, {277}, L55.

\ni Field, G.B. \& Rogers, R.D., 1993, ApJ, 403 94.

\ni Haardt F. \& Maraschi, L., 1993 MNRAS, 413, 507.

\ni Herrnstein, J.R. et al., ApJL in press, (preprint astro-ph/9802264).

\ni Ichimaru, S., 1977, Ap J, 214 840. 

\ni Frank, J., King, A, \& Raine, D., 1992, {\it Accretion Power in 
Astrophysics}, (Cambridge: Cambridge Univ. Press).

\ni Lasota, J.-P., Narayan, R. \& Yi, I1996, A\&A, 314 813.

\ni Mahadevan, R. 1997, ApJ, 477 585.

\ni Narayan R., \& Yi, I., 1995a, 428 L13.

\ni Narayan R., \& Yi, I., 1995b, 452 710.

%\ni Narayan, R., \& Yi, I,, \& Mahadevan, R., 1995, Nature {\bf 624}, 373.

\ni Narayan, R., Mahadevan, R. \& Quataert E., 1998a, 
in {\it Theory of Black Hole Accretion} M.A. Abramowicz, G. Bjornsson, \& J.E. Pringle eds. (Cambridge: Cambridge Univ. Press).

\ni Narayan, R., et al, 1998b, ApJ, 492, 554.

\ni Papaloizou, J.C.B. \& Lin, D.N.C., 1995, ARAA, 33, 505 

\noindent Parker, E.N., 1979, {\sl Cosmical Magnetic Fields}, (Oxford:
Clarendon Press) p 513.

\ni Pringle, J.E., 1981, ARAA, 19, 137

%\ni Rees, M.J., in {\it  Galactic Center}, G.R. Riegler \& R.D Blandford eds.,
%(New York: AIP), p166.

\ni Rees, M.J. et al, 1982, Nature 295, 17.

\ni Reid M.J., et al., 1989, ApJ, 336, 112.

\ni Reynolds, C.S. et al, 1998, MNRAS, 283 111.

\ni Shakura N.I., 1973, Sov. Ast., 16  5.

\ni Shakura N.I. \&  Sunyaev R.A., 1973, A.\&A., {24}, 337.

\ni Spencer R.E. \& Junor W., 1986, Nature, 321, 753.

\ni Stone J.M., et al., 1996, 453, 656.

\ni Xu, G. \& Chen X., 1997, ApJ 489, L29.

\end